# 5G Control Channel Design for Ultra-Reliable Low-Latency Communications

Hamidreza Shariatmadari, Sassan Iraji, Riku Jäntti (Aalto University)

Petar Popovski (Aalborg University)

Zexian Li, Mikko A. Uusitalo (Nokia Bell Labs)

*The fifth generation (5G) of wireless systems holds the promise of supporting a wide range of services with different communication requirements. Ultra-reliable low-latency communications (URLLC) is a generic service that enables mission-critical applications, such as industrial automation, augmented reality, and vehicular communications. URLLC has stringent requirements for reliability and latency of delivering both data and control information. In order to meet these requirements, the Third Generation Partnership Project (3GPP) has been introducing new features to the upcoming releases of the cellular system standards, namely releases 15 and beyond. This article reviews some of these features and introduces new enhancements for designing the control channels to efficiently support the URLLC. In particular, a flexible slot structure is presented as a solution to detect a failure in delivering the control information at an early stage, thereby allowing timely retransmission of the control information. Finally, some remaining challenges and envisioned research directions are discussed for shaping the 5G new radio (NR) as a unified wireless access technology for supporting different services.*

## Introduction

The fifth generation (5G) of wireless systems promises to offer new services for supporting a wide range of applications. According to the Third Generation Partnership Project (3GPP), main generic services for 5G include enhanced mobile broadband (eMBB), massive machine-type communications (mMTC), and ultra-reliable low-latency communications (URLLC) [1], [2]. eMBB targets high data rates, which were considered a common objective for previous generations of cellular systems. mMTC aims to provide connectivity for a large number of devices, which can further the development of the Internet of Things (IoT). URLLC is a communication service with strict requirements for availability, reliability, and latency [3].

URLLC enables mission-critical applications, such as industrial automation, augmented reality, and vehicular communications. The transmission links for these applications can be either one-to-one, one-to-many, or many-to-many. For instance, augmented reality and remote surgery applications require one-to-one communication links, while vehicular communications need one-to-one, one-to-many, and many-to-many links in order to provide connectivity among vehicles and road infrastructures.

The 3GPP considers two paths towards enabling the URLLC. The first path is based on the Long Term Evolution (LTE) and entails backward compatibility with the legacy LTE systems. The other path is based on the 5G new radio (NR) and compels forward compatibility with the 5G evolution. This paves the way for fundamental changes to the NR, which can bring better support for URLLC. While these two paths lead to different network designs, they might benefit from similar techniques for integrating URLLC [4].

However, URLLC can only be implemented if the high-reliability and low-latency features are addressed in the whole system [5]. The most challenging part is to meet these requirements in radio access networks (RANs). This is due to the dynamics of wireless channels. The RAN consists of physical channels that carry various types of information, generally categorized as data and control channels. These channels exhibit different impacts on the overall communication performance. Thus, different reliability and latency constraints are imposed to the channels according to the given communication service [4]. Since these constraints are usually stringent for URLLC, new approaches and designs are needed for the data and control channels.

This article presents some of the new features introduced in the upcoming releases of LTE and 5G NR that could be used to support URLLC. Then, reliability trade-offs between the data and control channels are described, which help to identify the reliability requirements for these channels. To meet the reliability constraints in the control channels, various solutions are presented that are potentially applicable in the design of 5G NR. Specifically, these solutions ensure high reliability for delivering scheduling request (SR), resource grant (RG), channel quality indicator (CQI) report, and hybrid automatic repeat request (HARQ) feedback. Furthermore, a flexible slot structure is proposed to identify a failure in delivering the control information at an early stage. This allows reducing the latency by taking the relevant actions timely.

## URLLC Requirements and Enablers

The target of 3GPP is to support a communication reliability corresponding to a block error rate (BLER) of $10^{-5}$ and up to 1 millisecond (ms) radio latency for delivering short packets up to 32 bytes. This target is specified by setting a user plane latency of 0.5 ms for uplink and downlink. The latency requirement is relaxed to 3-10 ms for supporting enhanced vehicle-to-everything (eV2X), which facilitates the autonomous driving, with larger packet sizes up to 300 bytes [1]. While these requirements are satisfactory for many mission-critical applications, more stringent requirements might be essential to support some other envisioned applications, particularly, in the realm of industrial automation and vehicular communications.

The 3GPP has introduced new techniques for LTE Rel. 14 and Rel.15 to support URLLC. These include fast uplink access, short transmission time interval (sTTI), and shortened processing time, thus reducing the user plane latency. In the legacy LTE, a user equipment (UE) needs to send an SR in order to be granted with the radio resources for transmitting its data. However, fast uplink access enables reserving radio resources for the UE, which can be utilized for uplink data transmissions whenever the UE has something to send. This reduces the latency as the UE does not need to send an SR and wait for the RG. Employing the sTTI is the other approach for reducing the transmission latency. The legacy LTE defines a subframe spanning over 14 symbols, resulting in a transmission time interval (TTI) of 1 ms. An sTTI can be formed by reducing the transmission duration, i.e., utilizing a mini-slot that is spanned over 2 to 7 symbols. The shortened processing time can further reduce the latency by sending the HARQ feedback faster than the legacy LTE, by which the feedback is sent after at least 4 subframes from the time of receiving the data. A potential enhancement for improving the reliability is the dual connectivity. In such a case, the UE can simultaneously communicate with multiple access nodes.

The 5G NR offers promising features that bring better support for URLLC. Some of the relevant features include access to the high bandwidths, support for massive multi-input multi-output (MIMO) antennas, enabling device-to-device (D2D) communications, introduction of new channel coding schemes, and configurable subcarrier spacing [2], [6]. The NR can access to a wide range of spectrum, including the millimeter wave (mmWave), which provides abundant radio resources for different services. In addition, employing the mmWave enables massive MIMO antenna systems, consisting of a large number of antennas accommodated at a base station, referred to as a gNB in 5G. This leads to better channel qualities and increase in the system capacity. The communication latency can be reduced by employing the D2D communications, in which UEs communicate directly without passing data through the gNB [5]. The NR supports both low density parity check (LDPC) and polar coding schemes. Specifically, LDPC is applied to both uplink and downlink data transmissions, which exhibits good BLER performance for URLLC. One of the nice features of the NR is its subcarrier spacing configurability with the values of 15, 30, 60, 120, and 240 kHz [6]. This allows accommodating different number of slots within a 1-ms-subframe and obtaining TTI of 1, 0.5, 0.25, 0.125, and 0.0625 ms, respectively. However, the highest subcarrier spacing that supports data transmissions is 120 kHz, corresponding to a TTI of 0.125 ms. In addition, a large variety of slot formats are introduced that bring high flexibility to the scheduling. The slot configurations can be categorized according to the symbol types, as illustrated in Figure 1. There are three different symbol types: uplink, downlink, and flexible. A UE shall assume downlink transmission through the downlink or flexible symbols, while it shall transmit by the uplink or flexible symbols [6]. The support of both downlink and uplink symbols within a slot is a promising feature for supporting URLLC, which allows reducing the latency. For instance, utilizing the slot format shown in Figure 1(h) for a downlink transmission enables the UE to receive the data at the beginning of the slot and to report the corresponding

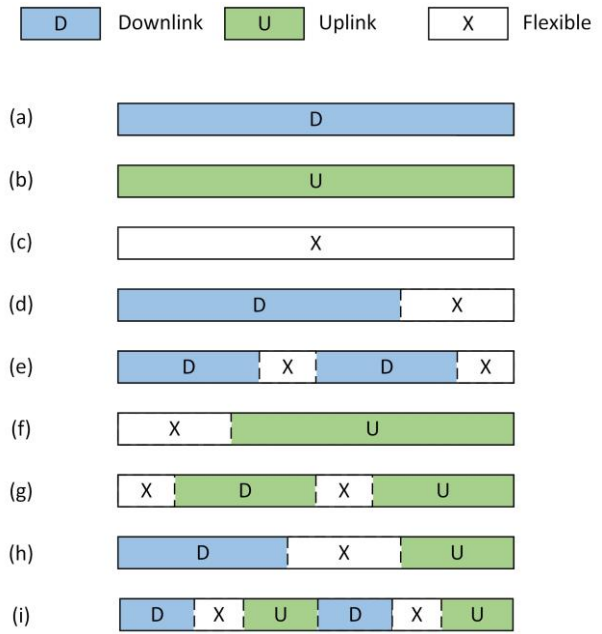

**Figure 1** *The illustration of slot formats in 5G NR.*

HARQ feedback at the end of the same slot. The same format can be utilized for an uplink transmission; the UE receives the uplink grant at the beginning of the slot and sends its data at the end of the slot.

URLLC entails providing reliable data and control channels. To understand better the effects of data and control channels on the overall communication reliability, we consider schedule-based communications for uplink and downlink data transmissions, as shown in Figure 2. For the uplink transmissions, a UE needs to send an SR to a gNB in order to access the radio resources. When the SR is detected, the gNB allocates the radio resources for the uplink data transmission. The gNB informs the UE about the allocated resources by sending a RG. The UE can transmit uplink data once the RG is decoded. If the gNB cannot retrieve the message correctly, it triggers the UE to retransmit the data. For adaptive data retransmissions, the gNB sends a new RG to the UE indicating the allocated radio resources for the data retransmission. The procedure of data retransmissions continues until either the message is decoded successfully or the maximum number of retransmissions is reached. The maximum number of retransmissions depends on the different parameters, such as latency requirement, TTI duration, and processing time. However, there is a common consensus that maximum number of retransmissions should not be more than one due to the latency constraint [1], [4].

For downlink transmissions, the gNB needs to know an estimate of the downlink channel quality for handling the link adaptation. This is done by using CQI report sent by the UE. Then, the gNB allocates radio resources for the downlink data transmission, according to the CQI report, and instructs the UE by sending the RG to monitor them for retrieving the message. Upon decoding the RG, the UE tries to decode the message and sends either an acknowledgement (ACK) or a negative-acknowledgement (NACK) signal to indicate the success or failure in the data reception. If the gNB does not receive an ACK signal, it retransmits the data. The gNB again instructs the UE to monitor the allocated resources for the data retransmission by sending a new RG. The procedure of data retransmissions continues until either the gNB finally receives an ACK signal or the maximum number of retransmissions is reached. Similar to the uplink transmissions, a maximum of one retransmission is envisioned due to the latency constraint.

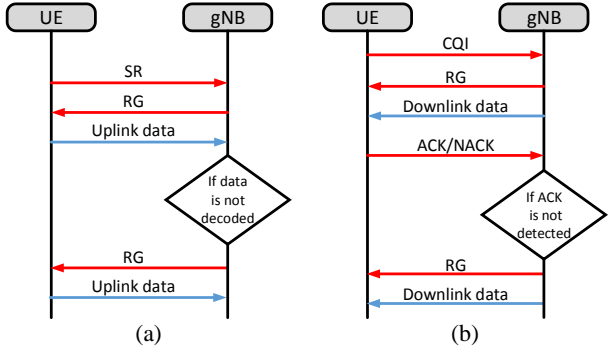

**Figure 2** *The schedule-based data transmissions in (a) uplink and (b) downlink.*

As explained, the uplink and downlink communications rely on transmitting data and control information. Both data and control channels are prone to the errors, affecting the overall communication reliability. However, the effects of the errors in data and control channels are different. For instance, one source of error is missing the RG that results in not sending the data in uplink or listening to the incoming downlink data. This error might happen during the initial transmission round and/or the retransmission round. In uplink, the gNB distinguishes this event when it does not receive any data from the UE, while in downlink, the gNB identifies this event when it does not receive an ACK nor a NACK signal, which is known as discontinuous transmission (DTX). In case the gNB identifies the missing of the RG for the initial transmission round, it can allocate more radio resources for the retransmission round in order to compensate the loss of initial transmission. However, there is a chance that the gNB detects the DTX erroneously as an ACK signal, then no retransmission is triggered. Another type of error is related to the CQI report, which carries an index that is derived according to the measured signal-to-interference-plus-noise ratio (SINR) and BLER target for the data transmission. The gNB might decode the CQI report wrongly as a higher or a lower value. Decoding the CQI report as a lower value results in employing an excessively robust modulation and coding scheme (MCS) for data transmission, thereby not degrading the communication reliability. However, incorrectly decoding the CQI report as a higher value leads to use of a MCS with a high transmission rate, which is less reliable. Another type of error is related to misinterpretation of ACK/NACK signals. The erroneous decoding an ACK as a NACK triggers unnecessary data retransmission, which results in wasting of resources. While, the erroneous decoding of a NACK as an ACK leads to absence of a necessary retransmission. Note that the errors of ACK/NACK signals affect only the retransmission round.

Let us consider uplink data transmissions. The failure rates of delivering the SR and the RG are $\epsilon_{SR}$ and $\epsilon_{RG}$, respectively. The initial data transmission is performed with the BLER of $P_1$. The BLER of decoding the message using the received information from the both initial data transmission and retransmission is $P_{1,2}$.

The BLER of $P_2$ is considered for decoding the message when the initial transmission is not triggered, due to missing the RG. Considering the errors of data and control channels, the success probability of delivering a message can be expressed as [4]

$P_{UL} = (1 - \epsilon_{SR})(1 - \epsilon_{RG})\{(1 - P_1) + P_1(1 - \epsilon_{RG})(1 - P_{1,2})\} + \epsilon_{SR}(1 - \epsilon_{SR})(1 - \epsilon_{RG})(1 - P_1) + (1 - \epsilon_{SR})\epsilon_{RG}(1 - \epsilon_{RG})(1 - P_2)$.

Figure 3(a) illustrates the reliability requirements for the control information to meet the reliability of $1 - 10^{-5}$ in uplink. The initial transmission is performed with three different reliabilities, while the retransmission ensures achieving the BLER of $10^{-5}$, i.e., $P_{1,2} = 10^{-5}$. It is assumed that $P_2 = P_1$. The target of communication reliability can be met only if the error rates of the control information are within the reliability regions. It can be observed that there are trade-offs between the reliabilities of data and control channels. For instance, $\epsilon_{SR}$ and $\epsilon_{RG}$ should be less than $10^{-4}$ if the initial data transmission ensures the BLER of 10%. These requirements can be relaxed by performing the initial transmission more reliably by using more robust MCS; however, this results in utilizing more radio resources for data transmissions [7], [8]. For example, the initial data transmission with the BLER of 1% entails that $\epsilon_{SR}$ and $\epsilon_{RG}$ be less than $10^{-3}$.

Now, we consider downlink transmissions and assume that the gNB has the perfect knowledge of the downlink channel quality. The failure rate of delivering the RG is $\epsilon_{RG}$. The initial transmission ensures the BLER of $P_1$. The probabilities of erroneously decoding a NACK as an ACK and a DTX are $\epsilon_{NA}$ and $\epsilon_{ND}$, respectively. While, the probabilities of incorrectly detecting a DTX as an ACK and a NACK are correspondingly $\epsilon_{DA}$ and $\epsilon_{DN}$. The BLER of decoding a message using the received information from the initial transmission and retransmission rounds is $P_{1,2}$. In case the gNB detects a DTX, it assumes that the UE could not receive any data information from the initial transmission round, hence, it can perform the retransmission more robustly. The BLER of decoding the message for this case is $P_{2D}$. However, in case the gNB decodes a DTX erroneously as a NACK, it retransmits data assuming that the UE has received the data from initial transmission round, although it cannot decode the message successfully. In this case, the BLER of decoding the message is reduced to $P_{2N}$. The success probability of delivering a message can be expressed as [4]

$P_{DL} = (1 - \epsilon_{RG})\{(1 - P_1) + P_1(1 - \epsilon_{NA} - \epsilon_{ND})(1 - P_{1,2}) + \epsilon_{ND}(1 - \epsilon_{RG})(1 - P_{2D})\} + \epsilon_{RG}(1 - \epsilon_{RG})\{\epsilon_{DN}(1 - P_{2N}) + (1 - \epsilon_{DN} - \epsilon_{DA})(1 - P_{2D})\}$.

Figure 3(b) illustrates the reliability requirements for the control information to achieve the reliability of $1 - 10^{-5}$ in downlink. The initial transmission round is performed with three different reliability targets. The data retransmission ensures the remaining BLER of $10^{-5}$, i.e., $P_{1,2} = P_{2D} = 10^{-5}$. In addition, it is assumed that $P_{2N} = P_1$. For the simplicity, we presume that $\epsilon_{A,N,D} = \epsilon_{NA} = \epsilon_{ND} = \epsilon_{DA} = \epsilon_{DN}$. The results show the similar trade-offs between the reliabilities of data and control channels. However, the reliability constraint for the HARQ feedback, i.e., $\epsilon_{A,N,D}$ is quite different from that for the RG. This is due to the fact that decoding the RG is prerequisite for both the initial transmission and retransmission rounds, while the ACK/NACK signals can only affect the retransmission round.

These observations indicate that URLLC entails higher reliability constraints for data and control channels than that offered by the legacy LTE (for instance, LTE complies 1%

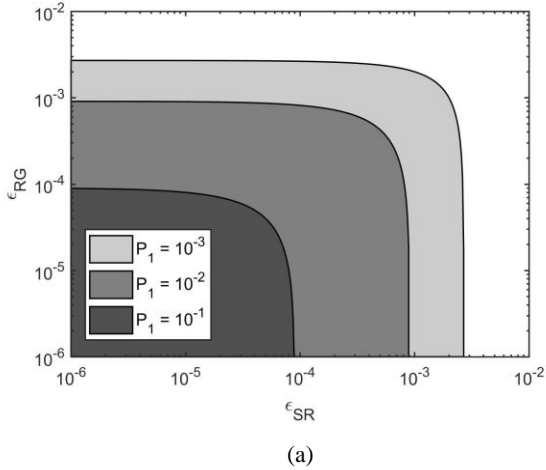

(a)

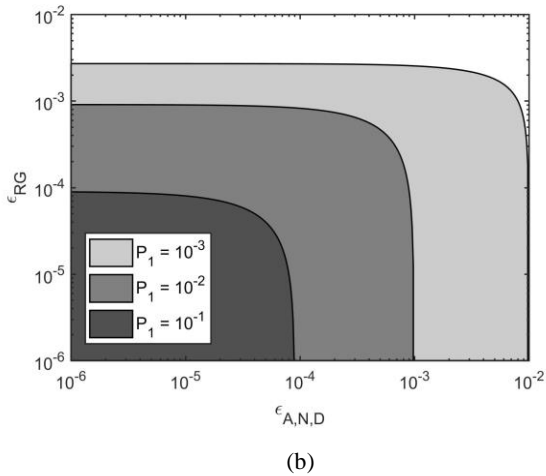

(b)

**Figure 3** *The reliability requirements for the control information in (a) uplink and (b) downlink.*

BLER for RG, 1% for the probability of ACK misdetection, and 1% BLER for CQI [9]). In the next section, we describe approaches that help in improving the reliability of control channels and offering better communication performance for supporting URLLC.

## Technical Challenges for Control Channels and Proposed Solutions

It was observed that the future cellular systems need to provide higher levels of reliability for data and control channels to support URLLC. While using redundant resources is a trivial solution for improving the reliability, it significantly reduces the communication efficiency. This motivates employing new approaches for designing the data and control channels to improve the reliability without degrading the communication efficiency. In addition, the new design should be able to support other services, such as eMBB and mMTC, at the same time. In the rest of this section, we present possible solutions for improving the reliability and the performance of delivering the control information. The promising solutions are provided separately for each type of control information.

### Scheduling request (SR)

A UE in a connected mode needs to send an SR to a gNB in order to be scheduled for uplink data transmission. In LTE, the SR is carried over the physical uplink control channel (PUCCH) and the base station uses energy detection to identify it. Each UE is configured with periodic orthogonal resources on PUCCH. The UE can send the SR only using predefined resources. When the UE wants to send data, it needs to wait until it has access to PUCCH. This introduces a random delay before the UE can access the channel. If the SR is not detected, the UE will not receive the RG for uplink transmission. Consequently, the UE needs to retransmit the SR, resulting in further delay. This delay can be reduced by assigning PUCCH resources to the UE more frequently, e.g., every TTI; nevertheless, this results in wasting a high portion of resources, particularly when the UE generates sporadic data traffic. In order to reduce the delay associated with the SR transmission while not wasting excessive radio resources, some of the following approaches can be considered:

- **Grant-free transmission:** Reserving radio resources for delivering the SR is not efficient for applications that generate sporadic data traffic. Instead, such applications can utilize grant-free transmission schemes to carry data without sending the SR. For instance, the UE can send data along with the preamble that is used for establishing a link [10]. However, the main issue with such schemes is the transmission collisions from different UEs that reduce the communication reliability. This can be improved by sending a few replicas of the message, which increases the chance of receiving one of them successfully.

- **Quality of service (QoS) based SR:** The SR in LTE does not carry any information about the constraints on the data delivery, in terms of the latency and reliability. In addition, the gNB does not know if the received SR is from the initial or the retransmission attempt. One enhancement is to include additional information regarding the communication requirements in the SR. For instance, the SR can carry information regarding the time budget and the required reliability for delivering the message. The gNB can utilize this information to allocate resources for transmission more efficiently. For instance, the gNB would select more robust MCS for the transmission if the time budget is low, due to the buffer latency or missing the previous SR by the gNB. It is shown that the inclusion of such information can also relax the reliability constraint on the SR [4].

- **Group-based SR:** The radio resources for the SR can be divided into different groups associated with different QoS. For instance, URLLC can access to a set of resources to send SR, while eMBB access to another set of resources. Users accessing the former resources are scheduled using shorter TTI compared to other users. This allows multiplexing different services more efficiently.

### Resource grant (RG)

The gNB delivers the downlink and uplink resource grants by sending the RG. In LTE, the RG is delivered over the physical downlink control channel (PDCCH). Decoding the RG is prerequisite for sending and receiving data, such that it requires high levels of reliability (see Figure 3). The following enhancements can be considered for delivering the RG:

- **Supporting higher aggregation levels:** LTE supports four different aggregation levels for PDCCH, which offer different reliability levels. For URLLC, the higher aggregation levels can be introduced to provide higher reliability. Another way is to send replicas of the RG using different resources in PDCCH. This allows exploiting the frequency diversity gain.
- **In-resource control signaling:** In order to provide more flexibility for encoding the RG, it can be carried over the data channel [11]. This allows employing different code rates for the RG. However, the UE needs to monitor a wide spectrum to find the RG, resulting in high power consumption.
- **Joint data and control channel coding**: The efficiency of coding scheme increases with the size of the input data [12]. However, the sizes of RG and data for URLLC are quite small, which reduce the communication efficiency. For downlink transmissions, the coding scheme can be applied jointly on the RG and the data in order to improve the efficiency. Nevertheless, this approach might increase the complexity of decoding procedure and the power consumption at the UE, as it needs to decode both the RG and the data.
- **Semi-persistent scheduling (SPS) and fast uplink access:** For periodic data transmission, a semi-persistent scheduling can be applied. In this way, the UE is informed about a set of resources that are reserved for it, such that the UE can send/receive data without the need to receive the RG. If the initial transmission fails, the gNB allocates additional resources and informs the UE by sending the RG [4]. The fast uplink access, which is introduced in the new releases of LTE, can be utilized for non-periodic data transmissions. This enables the UE to utilize the reserved resources only when it has data.
- **Advance (anticipative) RG transmission:** In LTE, the RG is sent for each data transmission or reception. In case a retransmission is required, a new RG is transmitted later. One of the solutions that is already agreed for 5G NR, is that the RG carries the resource allocations for a set of transmission/reception instances. For instance, the RG can indicate the radio resources for both the initial transmission and retransmission. This approach improves the reliability of RG detection, while imposing more signaling overhead as the RG carries information regarding the multiple transmissions.

### Channel quality indicator (CQI)

The CQI carries the downlink channel quality information. The UE derives the CQI according to the estimated SINR. The UE estimates the SINR by measuring the reference signals (RS) transmitted by the gNBs in different cells. The UE reports the CQI to the gNB, which is ultimately used for the link adaptation. In LTE, the UE maps the SINR to CQI by selecting the highest MCS that guarantees at least 10% BLER for a single transmission. In addition, there are altogether 16 CQI indexes that are represented by 4 bits. The CQI can be derived for the wideband, UE selected sub-bands, and the higher layer configured sub-bands. The wideband CQI is carried over the PUCCH, primarily using reserved radio resources periodically. In this case, the 4-bit CQI value is encoded into 20 bits for a protection against the noise and interference. Generally, there are two different issues associated with the CQI report. One is related to the CQI decoding, i.e., decoding a CQI as a higher or a lower value. Another issue for CQI report is the time gap between the channel measurement and the actual data transmission, during which the channel might change unfavorably [13]. Some of solutions for these issues are as follows:

- **Configurable CQI report**: Wideband CQI is carried over PUCCH using the same amount of resources. The lower coding rate can be utilized for CQI report in order to provide higher protection. This can be achieved by allocating more radio resources to the UE for reporting the CQI. Another way is to reduce the content of CQI report, e.g., using less than 4 bits to represents the CQI values. The cost is the lower performance of the link adaptation as only a subset of available MCS can be utilized.
- **Delay-based link adaptation:** The delay between the channel report and the data transmission degrades the accuracy of the CQI report. In order to obtain more accurate estimates of the channel quality, the UE can be configured to report PUCCH more frequently [5]. This would increase the signaling overhead and the power consumption. To compensate the effects of the outdated CQI report, the gNB can consider the CQI report delay while selecting the MCS for data transmissions. In this regard, a more robust MCS is selected when there is a long delay between the CQI report and downlink transmission [13]. This requires providing additional information for the scheduler, such as delay and channel variations.
- **HARQ feedback with an updated CQI:** To reduce the signaling overhead from the periodic CQI report, the UE can report an updated CQI after the initial downlink data transmission. For instance, the UE reports the CQI along with the NACK if the initial transmission fails.

### ACK/NACK signals

The UE needs to send either an ACK or a NACK signal after receiving the downlink data to indicate the success or failure in decoding the message. In LTE, these signals are carried over the PUCCH, using the same resource size for all the UEs. An erroneous detection of a NACK as an ACK signal results in suppressing the data retransmission, thereby degrading the overall communication reliability. However, the error in which an ACK is misinterpreted as a NACK results in unnecessary retransmissions of the data and thus wasting of radio resources. LTE has a 1% target for the ACK misdetection probability at a low SINR level with a single antenna. This reliability level is not sufficient for URLLC, as shown in Figure 3. The following approaches can improve the reliability of ACK/NACK signal detection.

- **ACK/NACK repetition:** In LTE, the ACK/NACK repetition is supported to improve the detection reliability for the UEs with bad channel conditions. The UE sends the same ACK/NACK signal multiple times over the consecutive TTIs. The gNB can configure the repetition factor. This scheme is similar to the TTI bundling that is used for physical uplink shared channel (PUSCH) in order to improve the reliability of data transmissions, particularly for the edge users. Although the ACK/NACK repetition improves the reliability of the detection, it introduces additional latency before the retransmission, because the retransmission starts only after all the ACK/NACK repetitions occur. To solve this issue, the ACK/NACK repetition can be performed during a single TTI while utilizing different frequency resources.

- **Asymmetric ACK/NACK signal detection:** As mentioned, protecting the NACK signal is more important than protecting the ACK signal, as erroneous NACK detection degrades the communication reliability [4], [7]. This brings forward the idea of using enhanced NACK protection by applying an asymmetric signal detection. For this purpose, the threshold for the binary hypothesis testing can be set in a way that the correct detection of NACK is favored. The cost of this approach is the higher rate of wrong detection of an ACK as a NACK compared to the case of employing a symmetric signal detection, in which the same probability is achieved for the miss detection of ACK and NACK. This results in performing more unnecessary retransmissions.
- **Early ACK/NACK transmission:** One of the issues in LTE is the high processing time for decoding the data. This postpones the ACK/NACK transmission to occur, i.e., at least 4 TTIs after receiving the data. This is due to the fact that ACK/NACK signal is transmitted after decoding the message. However, an early ACK/NACK transmission can be used by sending the ACK/NACK signal earlier based on the prediction of success or failure in decoding the message even before the message is decoded completely [14].
- **Multi-bit NACK:** LTE utilizes a single bit to carry ACK/NACK signals. Hence, the transmitter does not know how close the receiver's decoder was when attempting to retrieve the message upon receiving the NACK. For URLLC, this can result in significant decrease in communication efficiency, due to the limited number of transmission attempts. One effective solution is to utilize multi-bit NACK to adapt the redundancy of the data retransmission [15].

## Flexible slot Structure

One of the key challenges of URLLC is providing the high reliability for data transmissions with a limited number of transmission attempts, typically only one retransmission attempt is envisioned. This situation is aggravated when the errors occur in delivering the control information. For instance, a UE misses the transmission/reception chance if it cannot decode the RG successfully. This motivates us to exploit the flexibility of the 5G NR slot structures to detect a failure in delivering the control information and take immediate compensating actions. We propose a flexible structure scheme that is applicable to both time-division duplex (TDD) and frequency-division duplex (FDD). However, in this section we only focus on the TDD implementation as it is preferred widely due to the lower complexity and cost for UEs.

Figure 4(a) illustrates schedule-based uplink data transmissions in a TDD system. It is assumed that data should be delivered within two consecutive slots. Employed slots contain downlink, flexible, and uplink symbols. With the conventional approach of using a symbol either for uplink or downlink, the flexible symbols can be configured to carry uplink data. Accordingly, the gNB can deliver the downlink control information (DCI) that contains a RG at the beginning of each slot to instruct the UE in order to deliver uplink data. However, the UE misses the DCI in slot 1 and does not transmit uplink data. Hence, the gNB needs to send a new DCI in the next slot, which causes delay before the UE performs its first transmission. In addition, the gNB needs to allocate excessive radio resources for the data transmission in slot 2 as this is the last chance to deliver data within the time budget. In order to reduce this time gap, we propose to utilize the flexible symbols for both downlink and uplink transmissions. As shown in Figure 4(b), the gNB identifies that the UE has missed the RG as it does not transmit data in the uplink, i.e., DTX is detected. In this situation, the gNB retransmits the DCI using the flexible symbols. The UE decodes the retransmitted DCI and then starts transmitting data in the uplink. The retransmitted DCI can be the same as the initial DCI, or be different than that to allocate extended resources in the frequency domain for compensating the shortened transmission time. This approach gives the opportunity to have still two transmission attempts for delivering the data even if the DCI is missed.

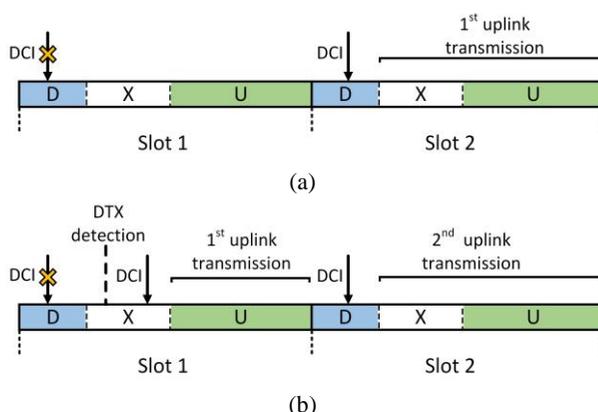

*Figure 4* Uplink data transmissions with an error in detecting the DCI utilizing, (a) the conventional slot structure and (b) the flexible

The proposed flexible slot structure can be also utilized for downlink data transmissions. As mentioned earlier, one source of errors is the use of an inappropriate MCS for delivering the data. The gNB might select an inappropriate MCS if it has decoded wrongly the CQI as a higher value or if the channel condition becomes worse drastically. In such conditions, there is a high chance that the UE cannot decode the message successfully. Figure 5(a) illustrates schedule-based downlink data transmissions with the conventional approach of using a symbol either for uplink or downlink transmissions. In this scenario, the flexible symbols are configured for downlink data transmissions. The gNB performs the initial downlink transmission over the slot 1 using an inappropriate MCS. The UE tries to decode the message after receiving the whole data and then sends the NACK signal along with the updated CQI for requesting the data retransmission. The gNB needs to retransmit the data using a more robust MCS. To address the issue of data transmission with an inappropriate MCS, we propose to utilize flexible symbols for both uplink and downlink transmissions, as shown in Figure 5(b). The UE decodes the DCI and determines the employed MCS and the resource allocations for the downlink transmissions. When the UE identifies that the employed MCS is not appropriate according to the current channel condition, it switches to the transmission mode immediately and sends an early NACK along with the updated CQI, using the resources allocated for its downlink transmission. When the gNB detects the early NACK signal, it terminates the concurrent data transmission and allocates new radio resources for the UE according to the updated CQI. The gNB sends a new DCI along with the data information in the same slot using a more robust MCS. As the refined downlink transmission uses a more robust MCS during a shorter time, the resource allocations should be expanded in frequency domain.

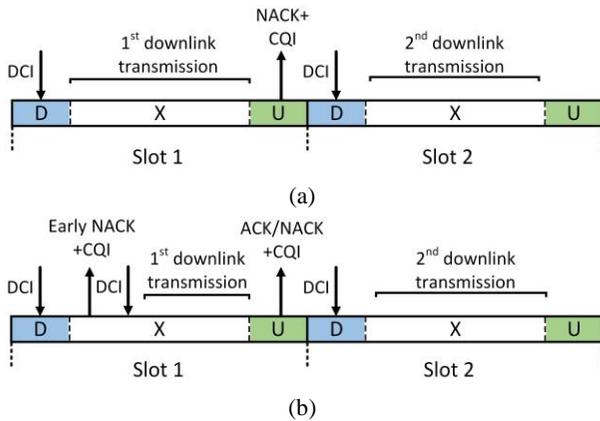

*Figure 5* Downlink data transmissions with an inappropriate MCS with (a) the conventional slot structure and (b) the flexible slot structure with an early NACK transmission.

It is observed that the proposed flexible slot structure, which can be implemented by using flexible symbols for both uplink and downlink transmissions, can reduce the latency and improve the communication efficiency. In order to employ the proposed scheme, the gNB should be able to operate in full-duplex mode to be able to send and receive simultaneously. However, the UE can still operate in half-duplex mode, which does not impose higher complexity in designing the UE radio.

## Conclusions and Future Directions

URLLC applications have different reliability and latency requirements. While the 5G NR has the potential to meet these requirements, it can benefit from nontrivial enhancements in order to bring better support for URLLC. This article presented solutions to improve the performance of delivering different control information, utilized for uplink and downlink transmissions. In addition, the proposed flexible slot structure allows detecting a failure in delivering the control information at an early stage and taking immediate compensating actions.

It was shown that data and control channels have different effects on the overall communication reliability. In addition, there are trade-offs between the reliability requirements for these channels. Hence, novel link adaptation and resource allocation schemes are required for the data and control channels. For instance, the resource allocations for the data channel should consider the reliabilities of control information, in addition to the link quality of the data channel. Another approach is to provide more flexibility for the control channels, hence, they can be configured to meet the communication requirements for different services. URLLC might be supported by both grant-based and grant-free transmission modes. The radio resources should be assigned for them optimally, and each user is configured to operate in one of these transmission modes according to its traffic type. For grant-based transmissions, the number of redundant transmissions, in time and frequency domains, is a key parameter that affects the communication reliability and efficiency. The redundant transmissions can be combined with specific patterns to provide a better performance. Another concern for the 5G NR is the multiplexing of different services, while satisfying their communication requirements. This can bring new challenges, particularly, when the system is faced by a sudden traffic surge from the URLLC users. One solution would be to puncture the radio resources that are allocated to other services in order to maintain the URLLC users. However, recovery mechanisms are also essential for allowing other services to resume their communications. In summary, these challenges should be taken into consideration to ensure efficient support of the URLLC in 5G systems.